\documentclass[pdflatex,sn-mathphys-num]{sn-jnl}

\usepackage{graphicx}%
\usepackage{multirow}%
\usepackage{amsmath,amssymb,amsfonts}%
\usepackage{amsthm}%
\usepackage{mathrsfs}%
\usepackage[title]{appendix}%
\usepackage{xcolor}%
\usepackage{textcomp}%
\usepackage{manyfoot}%
\usepackage{booktabs}%
\usepackage{algorithm}%
\usepackage{algorithmicx}%
\usepackage{algpseudocode}%
\usepackage{listings}%

\theoremstyle{thmstyleone}%
\theoremstyle{thmstyletwo}%

\theoremstyle{thmstylethree}%

\newcommand{\bcomment}[1]{\textcolor{blue}{\bf #1}}

\begin{document}

\title[How Fusion-Born Alpha Particles Suppress Microturbulence in Burning Plasmas]{How Fusion-Born Alpha Particles Suppress Microturbulence in Burning Plasmas}


\author*[1]{\fnm{Alessandro} \sur{Di Siena}}\email{alessandro.di.siena@ipp.mpg.de}

\author[1]{\fnm{Alejandro} \sur{Ba\~n\'on Navarro}}

\author[2]{\fnm{Pablo} \sur{Rodriguez-Fernandez}}

\author[2]{\fnm{Nathan T.} \sur{Howard}}

\author[1]{\fnm{Xin} \sur{Wang}}

\author[2]{\fnm{John} \sur{Wright}}

\author[2]{\fnm{Marco} \sur{Muraca}}

\author[3]{\fnm{Alexei} \sur{Polevoi}}

\author[1]{\fnm{Tobias} \sur{G\"orler}}

\author[1]{\fnm{Emanuele} \sur{Poli}}

\author[1]{\fnm{Roberto} \sur{Bilato}}

\author[3]{\fnm{Sun Hee} \sur{Kim}}

\author[3]{\fnm{Florian} \sur{Koechl}}

\author[2]{\fnm{Martin} \sur{Greenwald}}

\author[3]{\fnm{Alberto} \sur{Loarte}}

\author[1]{\fnm{Frank} \sur{Jenko}}

\affil*[1]{\orgdiv{Max Planck Institute for Plasma Physics}, \orgaddress{\street{Boltzmannstr. 2}, \city{Garching}, \postcode{85748}, \country{Germany}}}

\affil[2]{\orgdiv{MIT Plasma Science and Fusion Center}, \orgaddress{\street{167 Albany St}, \city{Cambridge}, \postcode{MA 02139}, \country{United States}}}


\affil[3]{\orgdiv{ITER Organization}, \orgaddress{\street{Route de Vinon sur Verdon CS 90 046 13067 St Paul lez Durance Cedex}, \country{France}}}

\abstract{A central unresolved question in fusion energy research is whether energetic alpha particles, the primary product of deuterium–tritium fusion reactions, help or hinder plasma sustainment. In a burning plasma — the operational regime of future devices such as ITER and SPARC — these particles become the dominant heat source, yet their effect on confinement has remained unknown. Here we report self-consistent simulations of burning plasmas that simultaneously evolve microturbulence, alpha-particle heating, and macroscopic profiles to steady state, finding that alpha particles substantially improve confinement. Fusion-born alpha particles weakly destabilize toroidal Alfv\'en eigenmodes (TAEs), which nonlinearly amplify zonal flows that shear apart and suppress ion-scale turbulence. The resulting reduction in heat loss drives core profile peaking, strengthening alpha heating by up to $25\%$ and closing a self-reinforcing loop. This feedback has no analogue in present-day devices, where external heating dominates, and reveals an intrinsic pathway toward improved confinement in burning-plasmas.}

\maketitle

\bmhead{Introduction}\label{sec1}

Achieving controlled magnetic confinement fusion would provide a virtually inexhaustible source of carbon-free energy. In magnetic confinement devices, this goal ultimately requires the realization of a burning plasma, in which fusion-born alpha particles from deuterium–tritium reactions become the dominant heat source sustaining the discharge. In this regime, external heating becomes secondary and the plasma is primarily self-heated, fundamentally altering its stability and transport properties \cite{Ongena_NP_2016,Shimada_2007}. Yet such conditions remain beyond the reach of present-day magnetic confinement experiments. Even the most advanced demonstration to date—JET’s deuterium–tritium campaign in 2024—produced fusion power that did not surpass the injected heating power \cite{Kappatou_2025}, highlighting the gap between current experiments and true burning-plasma conditions.

Whether alpha particles will reliably sustain a burning plasma, or instead trigger new pathways to confinement degradation, remains one of the most critical open questions for future fusion machines. In present-day devices, energetic alpha particles can resonantly excite shear Alfv\'en eigenmodes — electromagnetic oscillations that propagate along magnetic field lines — through wave–particle interactions \cite{Chen_RMP_2016}. These instabilities may redistribute energetic particles, flattening alpha pressure profiles and directly reducing heating efficiency \cite{Fasoli_NF_2007, Gorelenkov_NF_2014, Pinches_PoP_2015, Heidbrink_PoP_2008,Heidbrink2020}. At the same time, the coupling between Alfv\'enic activity and microturbulence may impact thermal transport, potentially improving confinement via turbulence suppression \cite{DiSiena_NF_2019, Mazzi_NatPhys_2022, Citrin_PPCF_2023, Na_NatRevPhys_2025, Garcia_NatComm_2024, Du_PRL_2026}, but also risking the opposite outcome through enhanced losses and profile redistribution \cite{Heidbrink_PoP_2008, Todo_2014, Biancalani_2021, Liu_PRL_2022}. Indeed, both experiments and numerical studies report regimes in which Alfv\'enic activity correlates with improved confinement, as well as cases where it leads to performance degradation.



What has been missing is a predictive framework that captures the full nonlinear coupling among alpha particles, turbulent transport, macroscopic profiles, and self-heating — all simultaneously and self-consistently. In a burning plasma these processes form a tightly coupled loop: changes in alpha confinement directly modify heating power, which alters temperature and density profiles, which in turn modifies turbulence-driven transport and the fusion reaction rate. This interplay cannot be approximated by decoupled or reduced models without losing the very physics of interest.



Here we report the first radially global, self-consistent gyrokinetic–transport simulations to evolve this coupled burning-plasma dynamics to steady state at fusion device scale. Applying this framework to reference scenarios for ITER \cite{Shimada_2007} and SPARC \cite{Creely2020} using the global gyrokinetic code \textsc{GENE} \cite{Jenko_PoP2000,Goerler_JCP2011} coupled to the transport solver \textsc{Tango} \cite{Parker_NF_2018,DiSiena_NF_2022}, we find that weakly destabilized toroidal Alfv\'en eigenmodes (TAEs) \cite{Cheng1986b,Fu1989} driven by alpha wave–particle resonances induce modest energetic-particle redistribution. However, they nonlinearly generate strong zonal flows — large-scale, toroidally symmetric $E \times B$  flows that shear apart and suppress turbulent eddies. These zonal flows regulate heat and particle transport, drive core profile peaking, and increase alpha self-heating by up to $25\%$ in SPARC and $18\%$ in ITER, establishing a self-reinforcing feedback loop that is intrinsic to burning-plasma operation.

\bmhead{Results}

\bmhead{Core confinement enhancement in burning-plasma simulations.}\label{sec2}




We quantify alpha-particle effects by comparing flux-matched \textsc{GENE}--\textsc{Tango}simulations run with and without alpha populations in \textsc{GENE}, while retaining alpha heating in the transport solver in both cases. The simulations self-consistently evolve turbulent heat and particle fluxes, alpha particles, heating sources, and plasma profiles ($T_e$, $T_i$, $n_e$) to steady state for SPARC and ITER reference scenarios. Alpha-particle profiles are updated iteratively via moments of a slowing-down distribution (see Method section).

\begin{figure}[h]
\centering
\includegraphics[width=0.9\textwidth]{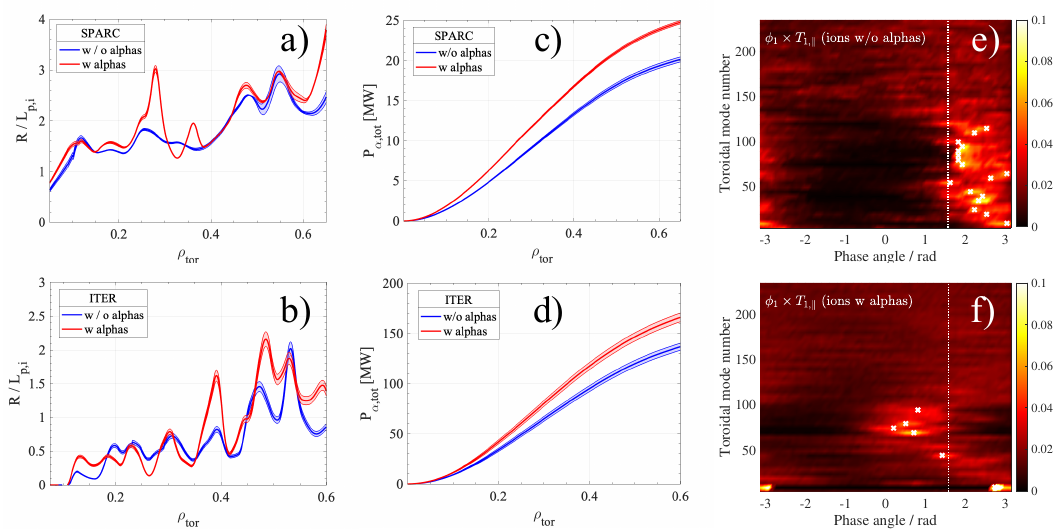}
\caption{\textbf{Core confinement and alpha heating enhancement in burning-plasma simulations}. Radial profiles of the steady-state logarithmic thermal ion pressure gradient ($R/L_{p,i}$) for SPARC a) and ITER b). A comparison between simulations with (red) and without (blue) alpha-particles reveals a pronounced gradient steepening, representing a major enhancement in thermal confinement. Radial profiles of the alpha-particle heating power $P_{\alpha}$ for SPARC c) and ITER d). These profiles illustrate the increased fusion heating power ($25\%$ in SPARC and $18\%$ in ITER) characteristic of the burning-plasma state when alpha-particle effects are self-consistently included in the global simulation. Cross-phase between temperature fluctuations ($T_{1,\parallel}$) and electrostatic potential ($\phi_{1}$) for SPARC at different toroidal mode numbers (n), simulated without e) and with f) alpha particles. The dotted white lines at $\pi/2$ denote the phase of maximum transport efficiency; the inclusion of alphas f) triggers a systematic shift toward zero, effectively minimizing the turbulence by reducing its heat-flux generation efficiency. }
\label{fig:fig1}
\end{figure}

Figure~\ref{fig:fig1}a), b) shows that including alpha particles produces a pronounced steepening of the logarithmic thermal ion pressure gradient at \( \rho_{\mathrm{tor}} \approx 0.3 \) in SPARC and \( \rho_{\mathrm{tor}} \approx 0.5 \) in ITER, indicating a strong local reduction of turbulent transport. Details of the SPARC and ITER scenarios are provided in the Methods section. The resulting enhancement of alpha heating power reaches $25\%$ in SPARC (figure~\ref{fig:fig1}c)) and $18\%$ in ITER (figure~\ref{fig:fig1}d)), with steady-state profiles converging within $1–2\%$ and corresponding heating uncertainties of $3\%$ and $\pm 5\%$ respectively.

\begin{figure}[h]
\centering
\includegraphics[width=0.9\textwidth]{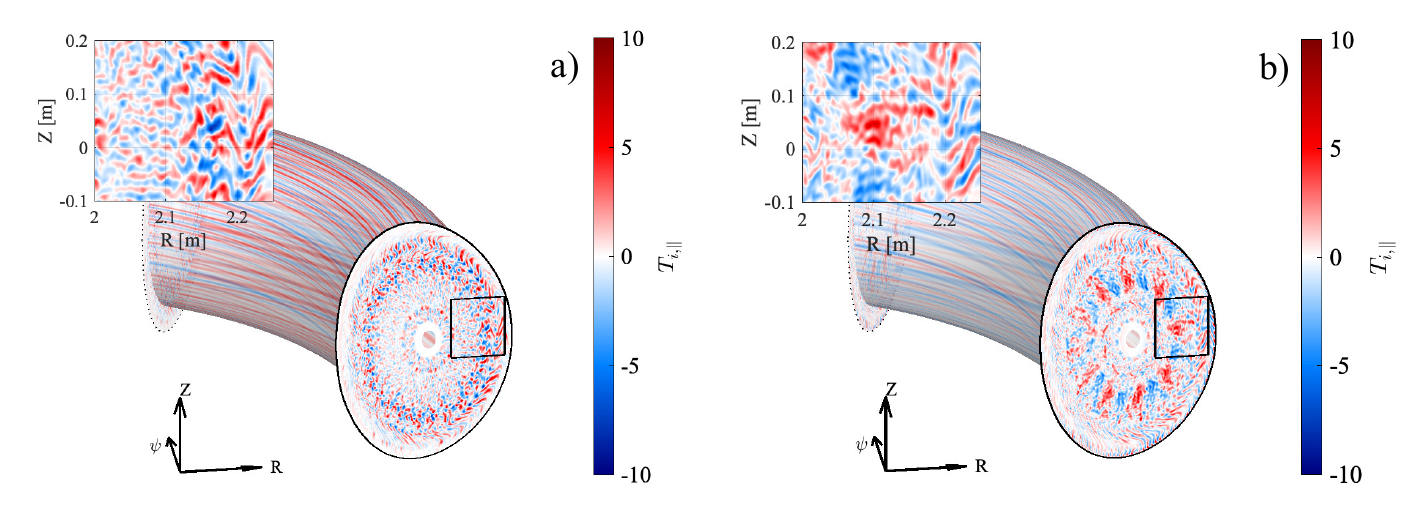}
\caption{\textbf{Alpha-particle-induced transition from microscopic to mesoscale turbulence}. Poloidal cross-sections of parallel temperature fluctuations $T_{1,\parallel}$ for SPARC global \textsc{GENE} simulations without a) and with b) alpha-particles. The inclusion of the alpha particles triggers a fundamental structural transition: from microscopic turbulent eddies a) to elongated, radially-extended mesoscale structures b).}
\label{fig:fig2}
\end{figure}
%


The suppression of turbulence is accompanied by a structural change in transport dynamics \cite{Ishizawa_2021}. Figure~\ref{fig:fig2} shows parallel temperature fluctuations transitioning from small microscopic eddies (on the scale of the ion Larmor radius, ranging from $\rho_i \sim 1.5\text{mm}$ in SPARC up to $\sim 4\text{mm}$ in ITER) to elongated, radially extended mesoscale structures (spanning tens of $\rho_i$) when alpha particles are included.

As turbulent structures spread across the poloidal cross-section, the cross-phases between temperature fluctuations and field perturbations undergo a systematic shift as shown in figure ~\ref{fig:fig1}e), f), from values near $\pi/2$, characteristic of the stronger turbulence regime without alpha particles, toward values approaching zero. Since the turbulent heat flux scales as $Q \propto |T_{1,\parallel}|\,|\phi_1|\,\sin\!\left(\alpha_{T_{1,\parallel},\phi_1}\right)$, where $Q$ is the heat flux, $|T_{1,\parallel}|$ and $|\phi_1|$ denote the amplitudes of the temperature and electrostatic potential fluctuations, and $\alpha_{T_{1,\parallel},\phi_1}$ is their cross-phase, this phase shift directly reduces transport efficiency, even with only minor changes in fluctuation amplitudes.



\bmhead{Alpha-driven destabilization of toroidal Alfv\'en Eigenmodes.}\label{sec3}

\begin{figure}[H]
\centering
\includegraphics[width=0.72\textwidth]{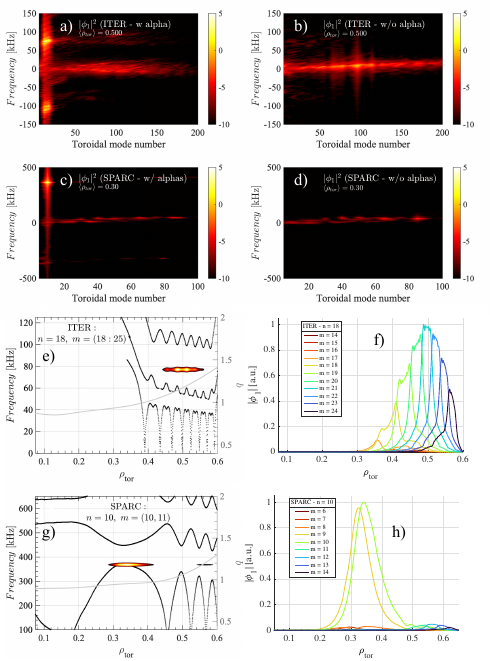}
\caption{\textbf{Alpha-driven TAE activity in ITER and SPARC.} Nonlinear spectrograms of the perturbed electrostatic potential at the location of the maximum of the alpha-particle flux for a), b) ITER and c), d) SPARC. A direct comparison between simulations with (a, c) and without (b, d) alpha-particle drive reveals the emergence of discrete, high-frequency instabilities. Radial profiles of the shear Alfv\'en continuum, computed with the ALCON code \cite{Deng2012b}, together with linear mode frequencies as a function of normalized toroidal flux ($\rho_{\mathrm{tor}}$), for ITER ($n=18$, e) and SPARC ($n=10$, g). Colored regions highlight the radial localization of the most unstable modes within the characteristic TAE gaps, while the safety-factor profile is overlaid in gray. Decomposition of the radial eigenmode structures into poloidal harmonics for ITER f) and SPARC h), illustrating the multi-mode coupling that defines the global TAE mode structure.}
\label{fig:fig3}
\end{figure}

The transition from microscopic to mesoscale turbulence arises from the destabilization of alpha-particle--driven Toroidal Alfv\'en Eigenmodes in the numerical simulations, which modify the turbulence characteristics and substantially affect the overall turbulent transport and steady-state plasma profiles in both SPARC and ITER. This is supported by spectrograms evaluated at the radial locations where the alpha-particle flux peaks in figure ~\ref{fig:fig3}a)-d). In simulations without alpha particles, no high-frequency activity is observed: the spectra remain dominated by fluctuations at frequencies comparable to linear drift-wave modes. Once alpha particles are included, clear high-frequency modes emerge at low toroidal mode numbers.

These modes lie at the center of the TAE gap and are driven unstable by the coupling of the poloidal harmonics $m$ and $m+1$. In SPARC, the TAE activity is localized around $\rho_{\mathrm{tor}} \approx 0.3$ (figure ~\ref{fig:fig3}g), h)), whereas in ITER the modes appear further out, around $\rho_{\mathrm{tor}} \approx 0.5$ (figure ~\ref{fig:fig3}e), f)). The ITER mode structure is also more complex, reflecting larger changes in the safety factor profiles around $\rho_{\mathrm{tor}} \approx 0.5$ which allows several poloidal harmonics to couple simultaneously at different radial locations. In SPARC, the near-zero magnetic shear at $\rho_{\mathrm{tor}} \approx 0.3$ leads to a simpler mode structure.


%

These modes are found to be only weakly unstable on the steady-state profiles, with growth rates smaller than those of the dominant drift-wave turbulence. Specifically, the ratio between the TAE growth rate and that of the dominant drift-wave instability is \( \gamma_{\mathrm{TAE}} / \gamma_{\mathrm{DW}} \approx 0.3\) in SPARC and \( \gamma_{\mathrm{TAE}} / \gamma_{\mathrm{DW}} \approx 0.2\) in ITER, where $\gamma_{\mathrm{DW}}$ is the dominant drift-wave growth rate. These values fall within the regime previously identified in present-day tokamaks where weakly unstable TAEs can beneficially modify turbulent transport rather than drive deleterious fast-ion--dominated turbulence \cite{Ishizawa_CommPhys_2025}. These TAEs induce a modest flattening of the alpha-particle pressure profile, corresponding to a reduction of approximately $\approx 7\%$ in SPARC and $\approx 9\%$ in ITER in the logarithmic pressure gradient.

Interestingly, the radial location of these modes closely coincides with the region where the thermal plasma profiles exhibit pronounced peaking in the flux-matching \textsc{GENE}--\textsc{Tango} simulations. This strong spatial correlation indicates that the destabilization of these modes, which occurs only when fusion-born alpha particles are included, plays a key role in suppressing thermal turbulence and enhancing confinement, as shown in figure \ref{fig:fig1}. 

\bmhead{Nonlinear Energy Transfer and Zonal Flow Amplification}\label{sec4}

\begin{figure}[h]
\centering
\includegraphics[width=0.9\textwidth]{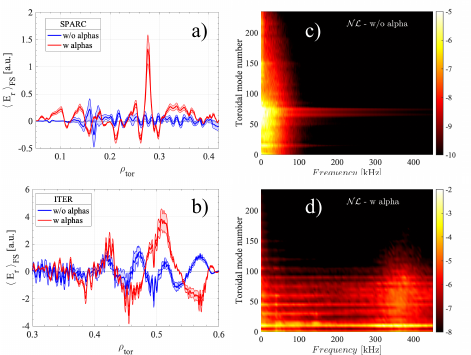}
\caption{\textbf{Nonlinear coupling and zonal structure generation by alpha-driven TAEs}. Radial profiles of the flux-surface averaged electric field $\langle E_r \rangle_{\mathrm{FS}}$ for SPARC a) and ITER b). The localized peaks in the alpha-driven cases (red) coincide with the regions of maximum TAE activity and pressure-gradient peaking. In c) and d) the radially-averaged spectral distribution of the nonlinear energy transfer ($\mathcal{NL}$) shown in a logarithmic color scale to the $n=0$ zonal component in SPARC as a function of toroidal mode number ($n$) and frequency. In the absence of alphas c), energy transfer is mediated by drift-wave turbulence at low frequencies. With alpha-particle drive d), the transfer is dominated by high-frequency TAE scales, exhibiting an amplitude increase of up to three orders of magnitude. This nonlinear redistribution of energy directly generates the shearing rates required for microturbulence suppression.}
\label{fig:fig4}
\end{figure}

The link between TAE destabilization and turbulence suppression is mediated by nonlinear mode–mode coupling. TAEs parametrically excite the n = 0 (toroidally symmetric) zonal component, producing localized radial electric fields and strong E×B shear \cite{Todo_2010,Qiu2016,DiSiena_NF_2019,Biancalani_2021,Liu2023,RuizRuiz_2025}. The resulting shear decorrelates turbulent structures, substantially reducing radial heat and particle transport \cite{Diamond_2005}, consistent with figure \ref{fig:fig1}. Figure \ref{fig:fig4}a), b) shows that this effect is radially localized and occurs only in simulations including alpha particles, precisely at the radial positions where the logarithmic pressure gradient increases. At these locations, the shearing rate $\omega_{E\times B}$ exceeds that of the most unstable linear mode by a factor of \(\sim 10\) in SPARC and \(\sim 3\) in ITER, placing the system well within the regime of strong shear suppression. The precise co-location of enhanced shear, pressure-gradient peaking, and TAE activity provides direct evidence of a causal link between alpha-driven Alfv\'enic dynamics and turbulence regulation.

To understand the underlying mechanism, we quantify the nonlinear energy transfer from finite toroidal mode numbers to the zonal component \cite{Navarro_2011,Nakata_2012}. Figure~\ref{fig:fig4}c), d) presents the nonlinear power transfer in the SPARC simulations without and with alpha particles. The nonlinear energy transfer for species $s$ is obtained from the free-energy balance as
\begin{align}
T_s(x,n) = \Re \Bigg[ \int dz\, d^3v\; \delta f_{s,n}^*(x,z,\mathbf{v}) 
\left( \frac{\partial \delta f_{s,n}}{\partial t} \right)_{\mathrm{NL}} (x,z,\mathbf{v}) \Bigg],
\end{align}
where $(\partial_t \delta f{s,n})_{\mathrm{NL}}$ denotes the nonlinear contribution to the time evolution of the perturbed distribution function for toroidal mode number $n$, $\Re{\cdot}$ indicates the real part, and $x,z,\mathbf{v}$ the coordinates in the phase space. This diagnostic quantifies the radially and spectrally resolved nonlinear redistribution of free energy.

In the absence of alpha particles, energy exchange with the zonal component is dominated by drift-wave turbulence at intermediate toroidal mode numbers $n \in [25,75]$, with transfer occurring at frequencies well within the drift-wave range. When alpha particles are included, a qualitatively different behavior emerges: the nonlinear transfer exhibits a strong modulation at the TAE frequencies and at the toroidal mode numbers where the TAEs are destabilized. At these scales, the amplitude of the nonlinear energy transfer to the zonal component increases by up to three orders of magnitude compared with the case without alpha particles.

These results demonstrate that the enhanced radial electric field (in absolute value) responsible for strong turbulence suppression---and the consequent increase in pressure gradients and alpha heating---is directly generated by alpha-driven TAEs through nonlinear coupling to zonal flows. The same mechanism is observed in the ITER simulations, establishing a consistent connection between TAE destabilization, zonal-flow amplification, profile peaking, and enhanced fusion gain across different fusion machines and burning-plasma conditions.

\bmhead{Discussion}\label{sec5}

\begin{figure}[h]
\centering
\includegraphics[width=0.9\textwidth]{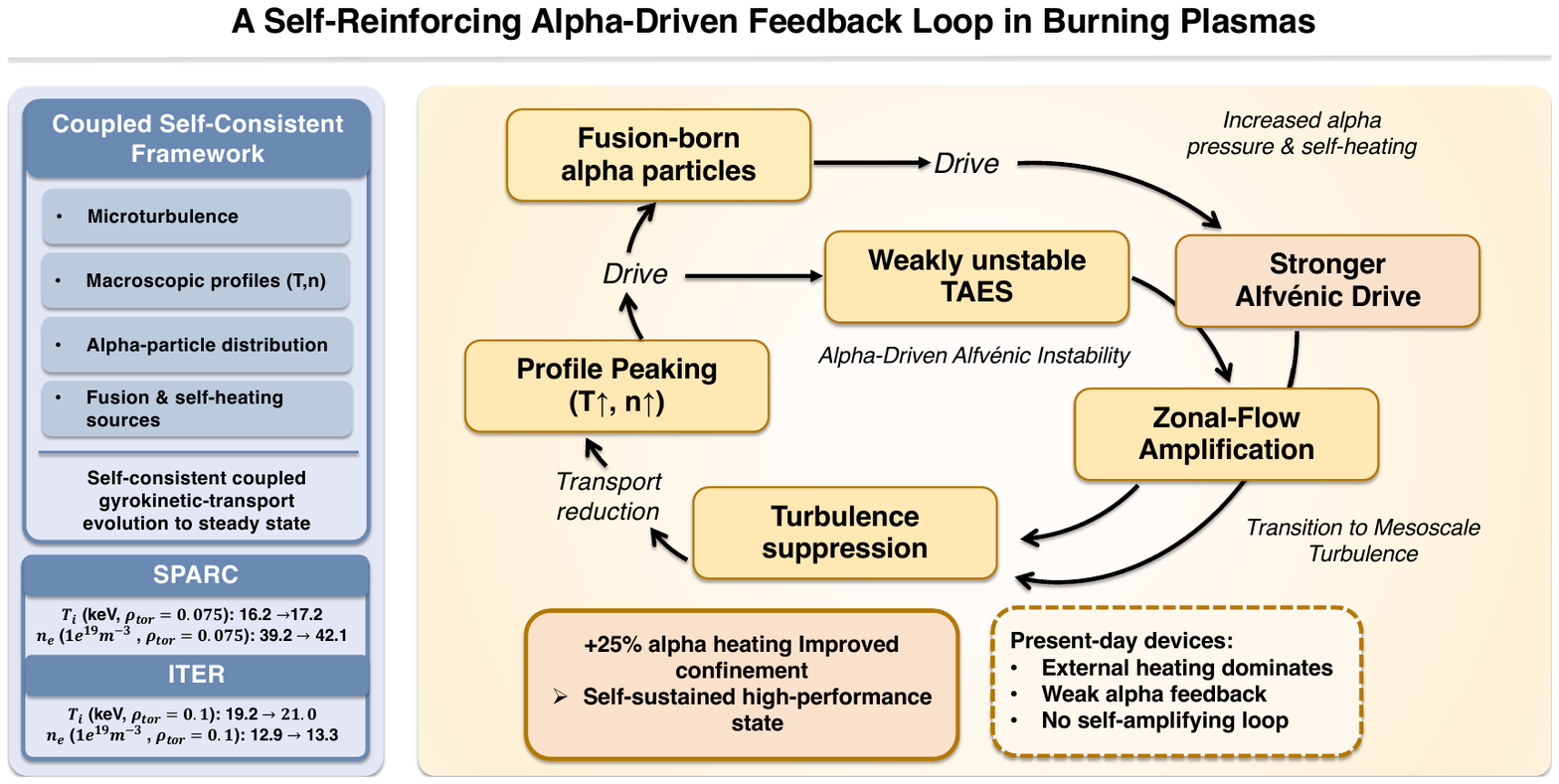}
\caption{\textbf{Schematic cartoon of turbulence regulation and self-heating in burning plasmas}. Fully self-consistent gyrokinetic–transport simulations reveal a nonlinear feedback mechanism in which fusion-born alpha particles drive weak toroidal Alfv\'en eigenmodes (TAEs), leading to zonal-flow amplification and mesoscale turbulence reorganization. The enhanced zonal shear suppresses microturbulence and reduces transport, enabling core profile peaking in temperature and density. The resulting increase in fusion power strengthens alpha-particle pressure and self-heating, further amplifying the Alfv\'enic drive and closing a positive feedback loop. Unlike present-day devices, where external heating dominates and alpha feedback is weak, this mechanism produces a self-sustained high-performance state characterized by up to $25\%$ increased alpha heating and improved confinement under burning plasma conditions.}
\label{fig:fig5}
\end{figure}

The simulations presented here reveal a degree of self-organization in burning plasmas that has not been observed or predicted in present-day devices. Rather than acting as a passive heat source, fusion-born alpha particles actively regulate turbulence through their coupling to Alfv\'enic modes and zonal flows, leading to a qualitatively different transport regime as illustrated in figure \ref{fig:fig5}. The self-reinforcing feedback loop where alpha heating drives TAEs, TAEs amplify zonal flows, zonal flows suppress turbulence, reduced transport raises core temperatures, higher temperatures increase alpha heating, operates only because external heating is no longer dominant. In ITER and SPARC, this mechanism is projected to raise alpha heating by $18–25\%$ above predictions that neglect alpha-particle dynamics, with direct implications for estimates of fusion gain Q.

For burning plasma operations, these results suggest that burning-plasma diagnostics sensitive to TAE activity — such as reflectometry and soft X-ray arrays — may provide early signatures of the onset of this beneficial feedback. Identifying the transition from drift-wave-dominated to TAE-mediated transport could serve as an operational marker for optimized performance. For compact high-field devices such as SPARC, the stronger TAE drive and sharper profile peaking predicted here indicate that the mechanism may be even more pronounced than in ITER, potentially contributing to fusion gains significantly above the $Q \approx 7–10$ range estimated by models that neglect alpha dynamics.

Several physical processes not captured in the present framework represent natural extensions. Large-scale MHD activity, in particular $n = 1$ internal kink and fishbone modes, can redistribute energetic particles and may additionally contribute to zonal flow amplification through nonlinear coupling. The self-consistent evolution of impurity species, including helium ash accumulation, is equally important as it directly influences plasma dilution and fusion reactivity. Fully resolving these effects would increase computational cost by up to an order of magnitude beyond current GPU-accelerated infrastructure. The robustness of the identified mechanism across the substantially different SPARC and ITER scenarios — differing by a factor of ~2 in magnetic field, major radius, and plasma current — nonetheless strengthens confidence that alpha-driven turbulence regulation is a fundamental feature of the burning-plasma regime rather than a scenario-specific artifact.

Ultimately, understanding and controlling these multi-scale interactions may prove essential for maximizing fusion device performance. The present results motivate dedicated experimental searches for the signatures of this mechanism, in particular, the co-location of weak TAE activity, elevated zonal flow shear, and core profile peaking, as ITER approaches burning-plasma conditions.

\bmhead{Methods}\label{sec7}

\bmhead{Burning-Plasma Scenarios in SPARC and ITER}

The SPARC primary reference discharge (PRD) considered in this work is a 50–50 deuterium–tritium H-mode plasma designed to operate in a high-field, compact configuration, with toroidal magnetic field \( B_t = 12.2\,\mathrm{T} \), major radius \( R \simeq 1.85\,\mathrm{m} \), and plasma current \( I_p = 8.7\,\mathrm{MA} \) \cite{Creely2020}. The scenario targets high confinement through auxiliary heating, consisting of \( \sim 11\,\mathrm{MW} \) of centrally deposited ICRH, in the absence of core particle refuelling and with negligible external torque. Previous studies have extensively investigated this operating point using a hierarchy of models, ranging from reduced turbulent transport models \cite{Muraca_2025} to local gyrokinetic simulations for core confinement \cite{Rodriguez_Fernandez_2022}, using an EPED-predicted peeling-limited pedestal \cite{Hughes_JPP_2020}, and neglecting the dynamical impact of fusion-born alpha particles, yielding predicted fusion gains in the range \( Q \approx 7\text{--}11 \). In parallel, hybrid MHD modelling has been employed to assess the stability of alpha-particle-driven modes \cite{Tinguely_2025}. Further details on the reference scenario and its modelling basis are provided in Ref.~\cite{Creely2020}.

For comparison, the ITER baseline scenario is modelled as a 50–50 deuterium–tritium H-mode plasma operating at lower magnetic field \( B_t = 5.3\,\mathrm{T} \) and larger machine size \( R \simeq \,6.2\mathrm{m} \), with higher plasma current \( I_p = \,15\mathrm{MA} \) and a target fusion gain of $Q \approx 10$ \cite{Mantica_PPCF_2019,Howard_NF_2025}.  In contrast to SPARC, ITER employs a combination of \( 33\,\mathrm{MW} \) of NBI and \( 20\,\mathrm{MW} \) of ECRH. While the NBI provides a source of external torque, the high energy of the beams ($1\,\mathrm{MeV}$) results in minimal core particle fueling, on the same order as the alpha-particle source itself and significantly lower than the fueling provided by edge pellets. The ECRH deposition is modeled as a Gaussian profile centered at $\rho_{\mathrm{tor}} = 0.4$, while NBI source profiles are computed using the PENCIL code \cite{Challis_NF_1989}. The pedestal density is prescribed at the pedestal top, and pellet fuelling is represented by a Gaussian source centred at \( \rho_{\mathrm{tor}} = 0.85 \). The pedestal temperature is adjusted such that the pedestal pressure matches the EPED-based expectation of \( \sim 130\,\mathrm{kPa} \) \cite{Polevoi_2015}. The toroidal rotation profile is obtained from QuaLiKiz \cite{Bourdelle_PoP_2007} calculations assuming a Prandtl number of unity. Additional details on the scenario setup can be found in Ref.~\cite{Mantica_PPCF_2019,Howard_NF_2025}.

\bmhead{Nonlinear gyrokinetic code \textsc{GENE} and simulation setup}

Nonlinear, radially global gyrokinetic simulations were performed using the GPU-accelerated \cite{Germaschewski_PoP_2021} Eulerian code \textsc{GENE}, which solves the five-dimensional Vlasov--Maxwell system for multiple kinetic species in arbitrary magnetic geometry. The simulations employ field-aligned coordinates \((x,y,z)\), where \(x\) denotes the radial coordinate, \(y\) the binormal direction, and \(z\) the coordinate along the magnetic field line. These coordinates are derived from magnetic flux coordinates \((\Psi, \chi, \varphi)\), with \(\Psi\) the poloidal flux, \(\chi\) the straight-field-line poloidal angle, and \(\varphi\) the toroidal angle. The radial coordinate is \(\rho_{\mathrm{tor}} = \sqrt{\Phi_{\mathrm{tor}}/\Phi_{\mathrm{LCFS}}}\), where \(\Phi_{\mathrm{tor}}\) is the toroidal flux, and \(\Phi_{\mathrm{LCFS}}\) the toroidal flux at the last closed flux surface. The binormal coordinate is \(y = \mathcal{C}_y (q\chi - \varphi)\), with \(q\) the safety factor, \(\mathcal{C}_y = x_0/q(x_0)\) a normalization constant, and \(x_0\) the domain center. The two velocity-space dimensions are the parallel velocity \(v_\parallel\) and the magnetic moment \(\mu\).  

Spatial discretization uses finite differences in \(x\) and \(z\), and spectral methods in \(y\) with binormal mode number \(k_y\). Boundary conditions are periodic in \(y\), twist-and-shift along \(z\), and Dirichlet with radial buffer zones (5\% of the domain) in \(x\) to damp fluctuations near the boundaries. A Krook relaxation rate of the order of a tenth of the maximum linear growth rate is used, allowing phase-space redistribution for alpha particles. Fourth-order hyperdiffusion suppresses small-scale spectral accumulation from electron-temperature-gradient modes.

The plasma is modeled with kinetic electrons, a single effective thermal ion species representing a 50-50 deuterium--tritium mixture, and fusion-born alpha particles. All species are modeled with Maxwellian equilibrium distributions. Impurities are accounted via a prescribed effective charge profile. Magnetic equilibria were obtained from EFIT for SPARC and CHEASE \cite{CHEASE_1996} for ITER. The minimum toroidal mode numbers considered were \(n=5\) (SPARC) and \(n=6\) (ITER), related to \(k_y\) via \(n = k_y \mathcal{C}_y\). To capture multiscale burning-plasma dynamics, simulations employed up to \(3 \times 1024 \times 96 \times 48 \times 48 \times 32\) grid points in \((\text{species},x,y,z,v_\parallel,\mu)\), totaling \(\sim 21.7\) billion points and requiring \(\sim 2.5\times10^5\) node-hours on GPU-accelerated HPC systems.

Flux-surface-averaged radial energy \(Q_s\) and particle \(\Gamma_s\) transport are computed as:
\begin{align}
Q_s &= \Big\langle \int \frac{1}{2} m_s v^2 \, \delta f_{1,s} \left( \mathbf{v}_{E \times B} \cdot \mathbf{\nabla} x \right) d^3v \Big\rangle, \\
\Gamma_s &= \Big\langle \int \delta f_{1,s} \left( \mathbf{v}_{E \times B} \cdot \mathbf{\nabla} x \right) d^3v \Big\rangle,
\end{align}
where \(\mathbf{x}=(x,y,z)\), \(\mathbf{v}=(v_\parallel,\mu)\), \(\langle \cdot \rangle\) denotes flux-surface averaging, and \(\delta f_{1,s}\) is the perturbed distribution function for species \(s\). The generalized \(E \times B\) drift is $\mathbf{v}_{E \times B} = \frac{c}{B_0^2} \, \mathbf{B}_0 \times \nabla \bar{\xi}$, $\quad \bar{\xi} = \bar{\phi} - \frac{v_\parallel}{c}\bar{A}_\parallel + \frac{\mu}{q_s} \bar{B}_\parallel$, with \(\bar{\phi}\), \(\bar{A}_\parallel\), and \(\bar{B}_\parallel\) gyro-averaged potentials, and \(q_s\) the particle charge.

\bmhead{Transport modelling with \textsc{Tango}}

Profile evolution and power balance calculations are performed using the transport solver \textsc{Tango}, which self-consistently evolves macroscopic plasma profiles by assuming a multiscale separation between background profiles and plasma microturbulence \cite{Parker_NF_2018} using the 1D transport equations:
\begin{align}
\frac{\partial n_s}{\partial t} + \frac{1}{V'} \frac{\partial}{\partial x} \left( V' \, \Gamma_s \right)
    = S_n , \\
\frac{3}{2}\frac{\partial p_s}{\partial t} + \frac{1}{V'} \frac{\partial}{\partial x} \left( V' \, Q_s \right)
    = S_{s},
\end{align}
with $V' = \mathrm{d}V/\mathrm{d}x$ the differential volume of the flux surfaces, $Q_s$ and $\Gamma_s$ the flux-surface-averaged heat and particle fluxes (taken from \textsc{GENE}in the present study), $S_n$ and $S_{s}$ the external particle and heat sources. The sources included in \textsc{Tango} are collisional energy exchange, alpha heating, and radiation losses—including bremsstrahlung, line radiation, and synchrotron emission—based on updated plasma profiles. These calculations incorporate multiple impurity species, specifically helium ash, beryllium, tungsten, neon, and alpha particles. External sources include Ohmic heating, Ion Cyclotron Resonance Heating (ICRH), Electron Cyclotron Resonance Heating (ECRH), and Neutral Beam Injection (NBI), along with NBI fueling and pellet injection for particle channels. The alpha-particle density and temperature profiles are updated iteratively for each transport step via the first and second moments of the slowing-down distribution function as detailed in Ref.~\cite{Estrada_2006}.

In the \textsc{GENE}--\textsc{Tango} framework, radial particle and heat fluxes are extracted from global gyrokinetic \textsc{GENE} simulations and used as inputs to the transport solver. This allows the evolution of flux-surface-averaged density and pressure profiles on macroscopic timescales. The coupling follows a LoDestro-type implicit scheme, where turbulent fluxes are decomposed into effective diffusive and convective contributions to ensure numerical stability and implicit convergence. Following each transport step, updated profiles are returned to the gyrokinetic code, which restarts from its previous turbulent state and runs until new statistically stationary fluxes are reached. This procedure is repeated until the turbulent fluxes match the source terms, yielding a steady-state solution. Final convergence is verified through standalone gyrokinetic simulations using the converged profiles to ensure consistency between the coupled and uncoupled fluxes.

\textsc{GENE}--\textsc{Tango} has been validated against experimental data from several devices. In ASDEX Upgrade discharges, the framework successfully modeled regimes ranging from moderate external heating to high-power scenarios where energetic particle effects are dominant \cite{DiSiena_NF_2022,Disiena_NF_2024}. The framework has also been applied to JET D-T plasmas, successfully modeling the highest-performance discharges from the DTE1 and DTE2 experimental campaigns with excellent agreement between simulations and measurements \cite{DiSiena_NF_2025}. Furthermore, the successful \textsc{GENE}--\textsc{Tango} validation has been extended to optimized stellarators across multiple W7-X scenarios \cite{Fernando_2025}.

\backmatter

\bmhead{Acknowledgements}

The authors would like to acknowledge insightful discussions with C. Bourdelle. The views and opinions expressed herein do not necessarily reflect those of the ITER organisation. This work was partially funded by Commonwealth Fusion Systems. Numerical simulations were performed at the Pitagora, Leonardo and Jupiter supercomputers.

\bmhead{Data availability}

\bmhead{Code availability}

\bmhead{Author contribution}


\end{document}